\def\half{\frac{1}{2}}
\def\ph{\phantom{-}}  
\def\mb#1{\mbox{\boldmath{$#1$}}}
\documentclass{iopart}\usepackage{cite}\usepackage{graphicx}
\begin{document}
\hspace*{2.5 in}CUQM-131, HEPHY-PUB 879/09\\
\markboth{R.~L.~Hall \& W.~Lucha}{Semirelativistic $N$-boson systems bound by attractive pair potentials}

\title{Semirelativistic $N$-boson systems bound by attractive pair potentials}

\author{Richard~L.~Hall}
\address{Department of Mathematics and Statistics, Concordia
University,\\ 1455 de Maisonneuve Boulevard West, Montr\'eal,
Qu\'ebec, Canada H3G 1M8\\ rhall@mathstat.concordia.ca}

\author{Wolfgang~Lucha}
\address{Institute for High Energy Physics, Austrian Academy
of Sciences,\\ Nikolsdorfergasse 18, A-1050 Vienna, Austria\\
wolfgang.lucha@oeaw.ac.at}

\begin{abstract}We establish bounds on the energy of a system of $N$ identical bosons
 bound by attractive pair potentials and obeying the semirelativistic Salpeter equation.  The lower bound is provided by a `reduction', with the aid of Jacobi relative coordinates, to a suitably scaled one-body Klein-Gordon problem. Complementary upper energy bounds are provided by means of a Gaussian trial function. Detailed results are presented for the exponential pair potential $V(r) = -v\exp(-r/a).$ 
\end{abstract}


\vskip 0.2in
\maketitle

\section{Introduction}We consider a system of $N$ identical bosons which interact by attractive central pair potentials $V(r_{ij})$.  There is no restriction in general concerning the number $d$ of spatial dimensions, but, for definiteness in the presentation, we shall assume $d=3.$ We suppose that the potential has the form $V(r) = vf(r/a),$ where $r = |{\mathbf r}|,$ $v >0$ is the coupling parameter, $f(r/a) < 0$ is the shape of the potential, and $a>0$ is a range parameter.  The Salpeter semirelativistic Hamiltonian for a system of $N$ identical bosons of mass $m$ and individual momenta $p_i$ is given by
\begin{equation}\label{eq1}
H = \sum_{i=1}^{N}\sqrt{p_i^2 + m^2} + \sum_{1=i<j}^{N}vf(r_{ij}/a),
\end{equation}
where $r_{ij} = |\mb{r}_i -\mb{r}_j|$ and $p_i = |\mb{p}_i|.$
Our principal goal is to find or estimate $E$, the lowest energy of this system. The ease with which the $N$-body problem can be posed in the Salpeter theory is the main reason for our adopting the semirelativistic approach.  For suitable potentials, the Hamiltonian is bounded below and to a certain extent the spectral theory mimics the corresponding Schr\"odinger theory \cite{BSE,SE,Lieb96}.

One path to a lower bound is to exploit the boson permutation symmetry of the exact normalized ground state $\Psi$ to effect a `reduction' generated by the equation $E = (\Psi, H\Psi) = (\Psi,{\mathcal H}\Psi),$ where ${\mathcal H}$ is the Hamiltonian for a specially scaled one-body problem.  A lower bound to $E$ is then given by the bottom of the spectrum of ${\mathcal H}.$  Such reductions have a long history which is reviewed briefly in Ref.~\cite{HL08}. Different reductions are possible and lead to a variety of lower bounds. For example, we obtain immediately from Eq.(\ref{eq1})
\begin{equation}\label{eq2}
E = (\Psi, H\Psi) = (\Psi, H_2\Psi),
\end{equation}
where
\begin{equation}\label{eq3}
H_2 = \frac{N}{2}\left[\sqrt{p_1^2 + m^2} + \sqrt{p_2^2 + m^2} + (N-1)V(r_{12})\right].
\end{equation}
The reduction is not yet complete since $H_2$ is the Hamiltonian for a scaled two-body problem.  If we adopt new coordinates $\mb{R} = \mb{r}_1+\mb{r}_2$ and $\mb{r} = \mb{r}_1-\mb{r}_2$ for this problem, we find $\mb{p}_1 = \mb{P}+\mb{p}$ and  $\mb{p}_2 = \mb{P}-\mb{p}$.  Thus if we write $\sqrt{p_i^2 + m^2} = |\mb{p}_i + m\mb{k}|$, $i = 1,2,$  where $\mb{k}$ is a unit vector orthogonal to $\mb{p}_1$ and $\mb{p}_2,$ then an application of the triangle inequality yields
$$ 2|\mb{p} + m\mb{k}| = |(\mb{p}-\mb{P} + m\mb{k}) + (\mb{p}+\mb{P} + m\mb{k})| \leq |-\mb{p}_2 + m\mb{k}|+ |\mb{p}_1 + m\mb{k}|.$$
Thus we have the spectral inequalities
\begin{equation}\label{eq4}
H \geq H_2 \geq {\mathcal H}_{N/2} = N\sqrt{p^2+m^2} + \gamma vf(r/a) \geq {\mathcal E}_{N/2},
\end{equation}
where $\gamma = \half N(N-1)$ and ${\mathcal E}_{N/2}$ is the bottom of the spectrum of ${\mathcal H}_{N/2}$ and provides a lower bound to $E$, the ground-state energy of $H.$ We  note that this use of the triangle inequality is equivalent to setting the centre-of-mass momentum $\mb{P}$ of the two-body Hamiltonian $H_2$ to zero.

 For boson systems there may be an energy advantage (leading to a higher lower bound) if, from the outset, Jacobi (orthogonal) relative coordinates can be accommodated by the reduction scheme. Intuitively, the use of relative coordinates allows us to remove the centre-of-mass kinetic energy of the whole system, rather than dropping the centre-of-mass energy only at the final stage, after the reduction to a scaled
two-body problem such as $H_2.$  One way of achieving this for the Salpeter problem involves squaring the reduced one-body Salpeter equation of the form $\sqrt{p^2+m^2}\psi = (E-V)\psi.$ This allows us to  invoke some remarkable expectation identities for Jacobi coordinates: these identities, which we shall refer to explicitly below, are proved in Appendix~A.

Since we use relative coordinates both for the upper and lower bounds, in section~2 we define the Jacobi coordinates $\{\mb{\rho}_i\}_{i=2}^N$  and note some key properties. An upper energy bound $E_U = (\Phi, H\Phi)$ is provided by an application of a normalized Gaussian trial function 
$\Phi$.  This latter calculation is greatly simplified by the boson permutation symmetry of $\Phi$ and also by the fact that this particular trial function is not only symmetric under the permutation of the relative coordinates but also factors in the form $\Phi = \prod_{i=2}^N \phi(\mb{\rho}_i)$. The upper bound is discussed in section~3.  In section~4 we introduce a variant of the Jacobi system suitable for the derivation of a lower bound expressed in terms of a Klein--Gordon one-particle problem. The proof assumes that this Klein--Gordon problem has a discrete eigenvalue that is either (i) positive or (ii) monotone increasing as a function of the mass.  In section~5 we consider the example of the attractive exponential potential.

\section{Relative coordinates}Since boson permutation symmetry is naturally expressed in terms of the individual-particle coordinates $\{\mb{r}_i\}_{i = 1}^N$, it might seem an unwelcome complication to introduce relative coordinates $\{\mb{\rho}_i\}_{i=2}^N.$ Their advantage arises 
when we consider the reduction $\langle H\rangle = \langle {\mathcal H}\rangle$, where ${\mathcal H}$ is a scaled one-body problem.  Jacobi relative coordinates may be defined with the aid of an orthogonal
matrix $B$ transforming old $\{\mb{r}_i\}$ to new
$\{\mb{\rho}_i\}$ coordinates by
$[\mb{\rho}]=B[\mb{r}].$ The first row of $B,$ with~all entries
$B_{1i}=1/\sqrt{N},$ defines a centre-of-mass variable
$\mb{\rho}_1,$ its second a pair distance
$\mb{\rho}_2=(\mb{r}_1-\mb{r}_2)/\sqrt{2},$ and the $k$th
row ($k\ge2$) first has $k-1$ entries $B_{ki}=1/\sqrt{k(k-1)},$ the $k$th
entry $B_{kk}=-\sqrt{(k-1)/k},$ and zero for all remaining entries. The
momenta $\{\mb{\pi}_i\}$ conjugate to the
$\{\mb{\rho}_i\}$ read $[\mb{\pi}]=(B^{-1})^{\rm
T}[\mb{p}]=B[\mb{p}].$ Now, for an $N$-boson problem with an attractive
potential $V(r),$ let $\Psi(\mb{\rho}_2,
\mbox{\boldmath{$\rho$}}_3,\dots,\mbox{\boldmath{$\rho$}}_N)$ be the exact normalized ground-state eigenfunction corresponding to the
lowest energy $E.$ Boson permutation symmetry is a powerful constraint that greatly
reduces the complexity of this problem. Although a non-Gaussian wave function
is not necessarily symmetric in the Jacobi coordinates, we do have the
remarkable $N$-representability expectation expressions from Appendix A:~for $i,j > 1$
\begin{eqnarray}
\left(\Psi,(\mbox{\boldmath{$\rho$}}_i\cdot\mbox{\boldmath{$\rho$}}_j)\Psi\right)&=\delta_{ij}\left(\Psi,\mbox{\boldmath{$\rho$}}_2^2\Psi\right),\\    \left(\Psi,(\mbox{\boldmath{$\pi$}}_i\cdot \mbox{\boldmath{$\pi$}}_j)\Psi\right)&=\delta_{ij}\left(\Psi,\mbox{\boldmath{$\pi$}}_2^2\Psi\right).
\label{eq6}
\end{eqnarray}

\section{A Gaussian upper bound} 
We now take advantage of the special properties of the Gaussian trial function $\Phi$ to effect an upper reduction of $\langle H\rangle$. In the  centre-of-mass frame of the $N$-body system, where $\mb{\pi}_1 = 0$, we have from the last row of the matrix $B$ 
\begin{equation}\label{eq7}
\mb{p}_N =  \frac{1}{\sqrt{N}}\mb{\pi}_1 -\sqrt{\lambda}\mb{\pi}_N = -\sqrt{\lambda}\mb{\pi}_N,\quad \lambda =\frac{N-1}{N}. 
\end{equation}
Meanwhile our trial wave function $\Phi$ is not only a boson function but it is also symmetric and factors in the relative coordinates. We have explicitly:
\begin{equation*}
\Phi(\mbox{\boldmath{$\rho$}}_2,\mbox{\boldmath{$\rho$}}_3,\dots,
\mbox{\boldmath{$\rho$}}_N)
=\prod\limits_{i = 2}^N \phi(\mb{\rho}_i), \quad \phi(r) = \left(\frac{1}{\pi\sigma^2}\right)^{\frac{3}{4}}\,e^{-\frac{1}{2}(r/\sigma)^2},
\end{equation*}
where $\sigma > 0$ is a variational scale parameter.
The boson permutation symmetry of the trial function allows us to write $E \leq E_g = \left(\Phi,H\Phi\right),$ where we have
 
$$
E_g = \left(\Phi, \left[N\sqrt{\lambda \mbox{\boldmath {$\pi$}}_2^2 + m^2} +\gamma V(|\sqrt{2}\mbox{\boldmath{$\rho$}}_2|)\right]\Phi\right).
$$
Setting $\mb{r} =\mb{r}_1-\mb{r}_2 =\sqrt{2}\mb{\rho}_2$ and $\mb{p}  = \mb{\pi}_2/\sqrt{2}$, we find $E \leq E_g = (\phi,{\mathcal H}_g\phi)$, where
\begin{equation}\nonumber
{\mathcal H}_g = N\sqrt{2\lambda p^2+m^2}+ \gamma V(r).
\end{equation}
We now introduce some new parameters which will allow us to express both the upper and lower energy bounds in compatible succinct forms.
With $V(r) = vf(r/a),$ $ \lambda = (N-1)/N$ and $\gamma = \half N(N-1)$, we now define
$$\mu = ma/(2\lambda)^{\half}\quad e = Ea/(2\gamma^{\half}),\quad e_g = E_ga/(2\gamma^{\half}),\quad \nu = \gamma^{\half} v a/2.$$
The Gaussian upper bound then becomes
\begin{equation}\label{eqhg}
 e \leq  e_g = \min_{\sigma > 0}\left(\phi, h_g \phi\right),\quad{\rm where}\quad h_g = \sqrt{p^2 + \mu^2} + \nu f(r).
\end{equation}

We note parenthetically here that, if the same scaling is used for the reduced Hamiltonian ${\mathcal H}_{N/2}$ of Eq.~(\ref{eq4}), the result $h_2$ is comparable with $h_g$ and is given by
\begin{equation}\label{eqh2}
h_2 = \sqrt{p^2/(2\lambda) + \mu^2} +\nu f(r),\quad \lambda = (N-1)/N.
\end{equation}
We shall use this later in the paper to show that the $N/2$ bound is weak: we shall consider
\begin{equation}\label{eqe2g}
e_{2g} = \min_{\sigma > 0}\left(\phi, h_2 \phi\right) \ge e_2,
\end{equation}
where $e_2$ is the bottom of the spectrum of $h_2$, and observe (in particular cases) that $ e_{2g} < e_k$, where $e_k$ is the Klein--Gordon lower bound that we shall derive in the next section; hence, for these problems, we know, without calculating $e_2$, that $e_2 < e_k$.

However, our main point here is that the upper energy bound can be written in terms of an optimized expectation (with respect to a  Gaussian wave function)  of a special one-particle Salpeter problem with Hamiltonian $h_g.$
The kinetic-energy
expectation value may be expressed in terms of the modified Bessel function of
the second kind $K_{1}(x)$ \cite{Abramowitz}. Evaluating the integral, for
convenience, in momentum space, yields
$$\left(\phi,\sqrt{p^2+\mu^2}\phi\right)=\frac{\mu}{\sqrt{2x}}I(x),$$
where $\sigma = \frac{\sqrt{2x}}{\mu}$ and the integral $I(x)$ is defined~by
$$I(x)=\frac{4}{\sqrt{\pi}}\int_{0}^{\infty}\sqrt{2x+t^2}\,e^{-t^2}t^2\,dt = \frac{2}{\sqrt{\pi}}\,x e^x K_1(x).$$
Meanwhile, the potential-energy expectation becomes 
$$\left(\phi,f(r)\phi\right) = J\left(\frac{\sqrt{2x}}{\mu}\right),$$
where the integral $J(y)$ is given by
$$J(y) = \frac{4}{\sqrt{\pi}}\,\int_{0}^{\infty}e^{-t^2}f(yt)t^2\,dt.$$
We now use $s = 1/\sigma = \frac{\mu}{\sqrt{2x}}$ as a variational parameter and we obtain
\begin{equation}\label{eg}
 e \leq  e_g = \min_{s > 0}\left[s\,I\left(\frac{\mu^2}{2s^2}\right) + \nu\,J\left(\frac{1}{s}\right)\right].
\end{equation} 

\section{The lower energy bound}

In order to effect a new reduction that incorporates $\mb{\pi}_1 = \mb{0}$, we first choose to use $\mb{ p}_{N}$ and $\mb{ p}_{N-1}$ and we obtain the reduction
$$E=(\Psi,H\Psi)=\left(\Psi,\left[\frac{N}{2}\sqrt{\mb{ p}_N^2+m^2}+\frac{N}{2}\sqrt{\mb{ p}_{N-1}^2+m^2}+\gamma
V(|\mb{ r}_{N-1}-\mb{ r}_N|)\right]\Psi\right),$$
where $\gamma=\frac{1}{2}N(N-1).$
We note that the last column of $B$ (which defines $\mb{\pi}_N$) is given by 
$$[b, b, b, \dots, b, -(N-1)b],\quad {\rm where}\quad b = 1/\sqrt{N(N-1)}.$$
Now we express the equation for $E$ in terms of new coordinates defined by the following relations
$$\mb{ r} = \mb{ r}_{N-1}-\mb{ r}_N = \alpha \mbox{\boldmath {$\rho$}}_{N}-\beta\mbox{\boldmath {$\rho$}}_{N-1}, \quad r = |\mb{ r}|,$$
$$\left[ \begin{array}{c}
\mb{ R}\\
\mb{ r}\end{array} \right] = 
\left[\begin{array}{cc}
\beta & \alpha\\
\alpha & -\beta \end{array}\right]
\left[ \begin{array}{c}
\mbox{\boldmath {$\rho$}}_N\\
\mbox{\boldmath {$\rho$}}_{N-1}\end{array} \right],
\quad 
\left[ \begin{array}{c}
\mb{ P}\\
\mb{ p}\end{array} \right] = 
\half\left[\begin{array}{cc}
\beta & \alpha\\
\alpha & -\beta \end{array}\right]
\left[ \begin{array}{c}
\mbox{\boldmath {$\pi$}}_N\\
\mbox{\boldmath {$\pi$}}_{N-1}\end{array} \right],$$
where 
$$\alpha = \sqrt{\frac{N}{N-1}}> 1,\quad\beta = \sqrt{\frac{N-2}{N-1}}< 1,\quad \delta = \sqrt{\frac{N-2}{N}} < 1.$$
Consequently,
$$\alpha^2 + \beta^2 = 2,\quad b^2 + \beta^2 = \frac{1}{\alpha^2} = \lambda = \frac{N-1}{N},$$
and
$$(N-1)b = \alpha^{-1},\quad Nb = \alpha,\quad \delta = \frac{\beta}{\alpha},
\quad 1+\delta^2 = 2\lambda.$$
We note that because of the boson permutation symmetry, and after we set the centre-of-mass momentum variable $\mb{\pi}_1 = \mb{0},$ we have generally that
$$\langle \kappa(|\mb{ p}_{N-1}|)\rangle = \langle \kappa(|\mb{ p}-\delta \mb{ P}|)\rangle = \langle \kappa(|\mb{ p}+\delta \mb{ P}|)\rangle = \langle \kappa(|\mb{ p}_{N}|)\rangle,$$
where $\kappa(p)$ is any appropriate kinetic-energy function.
The expression for the energy now has the form $E = \left\langle{\mathcal H}\right\rangle,$ where
\begin{equation}\label{eq11}
{\mathcal H} = N\sqrt{\left(\mb{ p}+\delta \mb{ P}\right)^2 + m^2}+\gamma V(r).
\end{equation}
The Hamiltonian ${\mathcal H}$ is bounded below by the simple bound ${\mathcal H}_{N/2}$ of Eq.~(\ref{eq4}). This result is a consequence of the inequality
$$ 2\langle|\mb{p} + m\mb{k}|\rangle = \langle|(\mb{p}-\delta\mb{P} + m\mb{k}) + (\mb{p}+\delta\mb{P} + m\mb{k})|\rangle \leq  2\langle|\mb{p}+\delta\mb{P} + m\mb{k}|\rangle.$$
 We now consider the eigenequation ${\mathcal H}\psi = {\mathcal E}\psi,$ where $\psi(\mb{ r},\mb{ R}) \in {\mathcal D} \subset L^2(R^6)$ retains some of the boson permutation symmetry implications of the full wave function, namely from Eq.~(\ref{eq6}), $\langle \mb{ p}^2\rangle = \langle \mb{ P}^2\rangle,$
 $\langle \mb{ p}\cdot\mb{ P}\rangle = 0,$ and hence
\begin{equation}\label{eq12}
\langle (\mb{ p}+\delta \mb{ P})^2\rangle = \langle p^2 + \delta^2 P^2\rangle = (1+\delta^2)\langle p^2\rangle = 2\lambda \langle p^2\rangle.
\end{equation}
Let ${\mathcal E}$ be the minimum of $(\psi,{\mathcal H}\psi)$ corresponding to normalized
$\psi$ satisfying (\ref{eq12}) and let $\Psi$ be the exact normalized $N$-body wave
function, then
$$E=(\Psi,H\Psi)=(\Psi,{\mathcal H}\Psi)\ge(\psi,{\mathcal H}\psi)={\mathcal E}.$$
The eigenvalue equation for ${\mathcal E}$ reads
\begin{equation}\label{sala}
N\sqrt{(\mb{ p}+\delta \mb{ P})^2 +m^2}~\psi=\left({\mathcal E}
-\gamma v f(r/a)\right)\psi.
\end{equation}
Squaring these equal vectors and use of (\ref{eq12}), and also the scaling change (to new dimensionless variables) $\{\mb{r}, \mb{p}\}\rightarrow \{\mb{r}a, \mb{p}/a\}$,  implies
\begin{eqnarray}\label{salai}
\nonumber({\mathcal E}a)^2-N^2(ma)^2 & \\
\ge 4\gamma\inf_{\psi\in{\mathcal
D}}\left(\psi,\left( p^2+\frac{(va)({\mathcal E}a)}{2}f(r)-\frac{\gamma
(va)^2}{4}f^2(r)\right)\psi\right).
\end{eqnarray}

\begin{figure}[htbp]\centering\includegraphics[width=12cm]{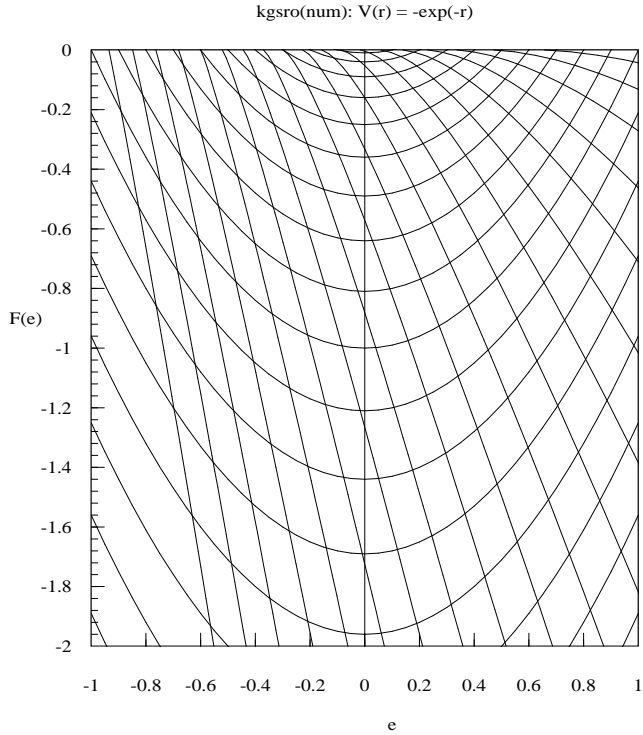}
\caption{The curves of Schr\"odinger energies $F(e)$ for $v = 1\, (0.25)\, 5.5$, and $e^2 - m^2$ for $ m = 0.1\, (0.1)\, 2.1$ for the exponential pair potential $V(r) = -ve^{-r}.$ The intersection points yield the Klein--Gordon ground-state energies $e = e(m,v).$ For these solutions to exist, the Schr\"odinger curves must be steeper than the $e^2 -m^2$ curves: thus we have $F'(e) < 2e$ at the intersection points, even when $e$ is negative; this is a sufficient condition for the Klein--Gordon energy $e$ to be a lower bound to the corresponding Salpeter energy $E>e$.
}\label{Fig:F(e)}\end{figure}

In contrast to our upper bound, where we worked directly with the Salpeter Hamiltonian (\ref{eqhg}), for the lower bound we are led to consider a complementary lower Klein--Gordon operator.  If the minimization on the right-hand side of (\ref{salai}) could  be effected exactly (as it can for the gravitational problem $f(r) = -1/r$ \cite{HL06}), then it would remain to extract a lower bound to ${\mathcal E}$ from the resulting inequality. For the general problem that we are considering here, we have to find another approach.  To this end we have explored what happens if we set ${\mathcal E} = {\mathcal E}_k$ in Eq.~(\ref{salai}), and solve the resulting Klein--Gordon equation for the ground-state energy $e.$ This Klein--Gordon equation is given explicitly by
\begin{eqnarray}\label{kg}
\nonumber\left(({\mathcal E}_ka)^2-N^2(ma)^2\right)\phi & \\
= 4\gamma\left(p^2+\frac{(va)({\mathcal E}_k a)}{2}f(r)-\frac{\gamma(va)^2}{4}f^2(r)\right)\phi.
\end{eqnarray}
This exercise is only useful if an {\it a priori} relation can be established between ${\mathcal E}_k$ and ${\mathcal E}$. More particularly, we shall now establish a relationship between ${\mathcal E}_k$ and the lower bound to ${\mathcal E}$ provided by the inequality given in Eq.~(\ref{salai}).  We employ a similar parameter set to that for the upper bound, that is to say $ \lambda = (N-1)/N$ and $\gamma = \half N(N-1)$, and also
\begin{equation}\label{params}
\mu = ma/(2\lambda)^{\half}\quad e_k = {\mathcal E}_k a/(2\gamma^{\half}),\quad \nu = \gamma^{\half} v a/2.
\end{equation}
We thus arrive at the Klein--Gordon equation in a standard form, namely
\begin{equation}\label{kg1}
(p^2 + \mu^2)\phi = \left(\nu f(r) -  e_k\right)^2\phi,
\end{equation}
and, equivalently, in Schr\"odinger form
\begin{equation}\label{kgs}
\left(p^2 + 2e_k\nu f(r) - (\nu f(r))^2\right)\phi = F(e_k)\phi = (e_k^2 - \mu^2)\phi.
\end{equation}
The spectral function $F(e_k)$ in Eq.~(\ref{kgs}) describes, for each fixed coupling $\nu$, how the lowest eigenvalue of the Schr\"odinger operator 
$p^2 + 2e_k\nu f(r) - (\nu f(r))^2$ depends on the energy parameter $e_k$; the Klein--Gordon eigenvalue exists if there is a solution of the equation $F(e_k) = e_k^2 - m^2;$ these curves are shown in Fig.~\ref{Fig:F(e)} for the case of an attractive exponential potential. 
Similarly, the original lower-bound inequality (\ref{salai}) becomes an equality in terms of the parameters (\ref{params}) and also a new mass parameter $\bar{\mu}\ge\mu$ used here as a `slack variable'. Thus, we have:
\begin{equation}\label{eqslack}
\left(p^2 + 2e\nu f(r) - (\nu f(r))^2\right)\phi = F(e)\phi = (e^2 - \bar{\mu}^2)\phi.
\end{equation}
In earlier work \cite{HL09} on the semirelativistic and Klein--Gordon {\it one}-body problems, we have proved that, under appropriate assumptions, the Klein--Gordon energy is  monotone in the mass. More specifically, in our application, we can state this result in terms of Eq.~(\ref{eqslack}), namely, for each fixed $\nu,$ if either $e > 0$, or $e \le 0$ and $e > \half F'(e),$ then $e'(\bar{\mu}) > 0.$  Thus, under these conditions (which must be verified for each potential), we have from Eqs.~(\ref{kgs}) and (\ref{eqslack}) that $e \ge e_k.$ Thus ${\mathcal E}_k$ provides us with a lower bound to the $N$-body ground-state energy $E$. In terms of the dimensionless energy parameters, our principal results are summarized by the following inequalities:
\begin{equation}\label{ebounds}
 e_k(\mu,\nu) \leq  e \leq  e_g(\mu,\nu),
\end{equation}
where $ e_k$ is given by solving the Klein--Gordon equation (\ref{kgs}), $ e_g $ is given by the variational equation (\ref{eg}), and the parameters are defined in equation (\ref{params}).  Thus, whenever a solution exists, we necessarily have $F'(e) < 2e.$ For example, in the case of the exponential pair potential we show this relationship explicitly in Fig. \ref{Fig:F(e)}.

\section{The exponential potential}
\begin{figure}[htbp]\centering\includegraphics[width=12cm]{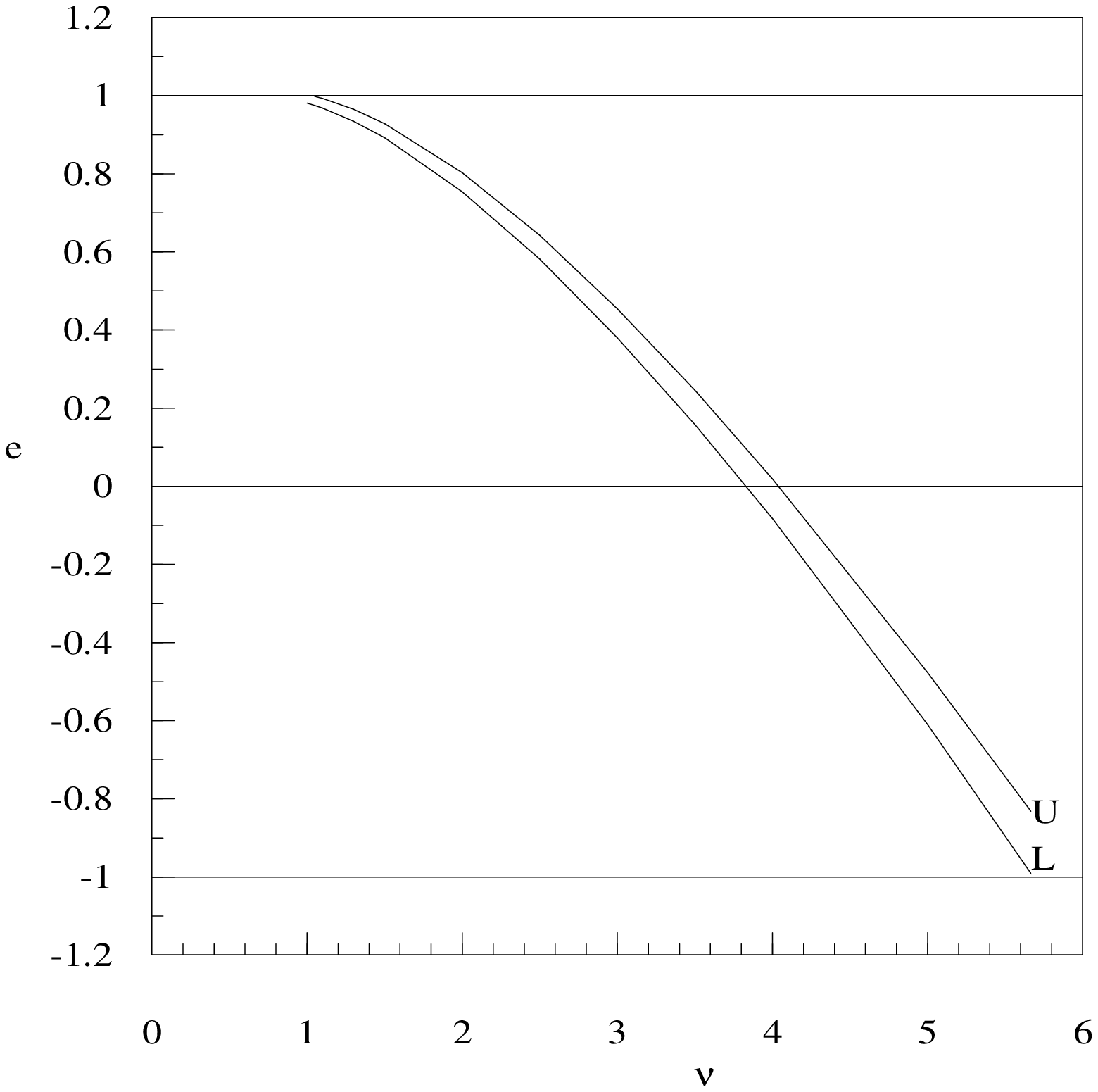}
\caption{Lower (L) and upper (U) energy bounds to the ground-state energy $e$ of the $N$-boson system with pair-potential shape $f(r) = -\exp(-r)$.
Dimensionless units $\mu = ma/(2\lambda)^{\half} = 1,$ $e = E a/(2\gamma^{\half}),$ $\nu = \gamma^{\half} v a/2,$ are used, in which 
$\lambda = (N-1)/N,$ and $\gamma = \half N(N-1)$. The lower and  upper data are given, respectively, by Eqs.~(\ref{ekexp}) and (\ref{egexp}).
}\label{Fig:E(v)}\end{figure}

We now consider the example with potential shape in dimensionless form given by $f(r) = -e^{-r}$.  The lower bound $e_k(\mu, \nu)$ is given by solving the explicit one-particle Klein--Gordon equation 
\begin{equation}\label{ekexp} 
(p^2 + \mu^2)\psi = (e_k + \nu e^{-r})^2\psi
\end{equation}
and the  upper bound $e_g(\mu, \nu)$ is given by
\begin{equation}\label{egexp}
e_g = \min_{s > 0}\left[s\,I\left(\frac{\mu^2}{2s^2}\right) + \nu\,J\left(\frac{1}{s}\right)\right],
\end{equation}
where
\[
I(x)=\frac{4}{\sqrt{\pi}}\int_{0}^{\infty}\sqrt{2x+t^2}\,e^{-t^2}\,t^2\,dt = \frac{2}{\sqrt{\pi}}\,x\,e^x\,K_1(x),
 \]
and
\begin{eqnarray*}
J(y) &= - \frac{4}{\sqrt{\pi}}\,\int_{0}^{\infty}e^{-t^2}e^{-yt}t^2\,dt \\
&= \frac{y}{\pi^{\half}} -\half(2+y^2)(1-{\rm erf}(y/2))\,e^{y^2/4}.
\end{eqnarray*}
For the choice $\mu = 1$ we exhibit some numerical results in Table~1 and Fig.~\ref{Fig:E(v)}.
\begin{table}[b]
\begin{center}
\caption{Energy bounds $e_k \leq e \leq e_g$ for the $N$-boson system bound by an attractive pair potential with shape $f(r) = -\exp(-r)$.
The parameters are given by $\mu = ma/(2\lambda)^{\half} = 1,$ $e = E a/(2\gamma^{\half}),$ $\nu = \gamma^{\half} v a/2,$ 
$\lambda = (N-1)/N,$ and $\gamma = \half N(N-1).$  The energy $e_{2g}$ is an upper bound to the $N/2$ lower bound $e_2$. 
 The table shows $e_{2g}$ for $N\rightarrow \infty$. Since $e_{2g} < e_k,$ we know that the Klein--Gordon bound $e_k$ is better than the $N/2$ bound, that is to say, for large $N$, $e_2 < e_{2g}< e_k.$}\medskip
\label{tab:Table 1}
\begin{tabular}{|l|l|l|l|} 
\hline
 $\nu$ & $e_{2g}$ & $e_k$ & $e_g$\\ \hline
 $0.6$ & $\ph 0.993207$ & & \\ \hline
 $0.8$ & $\ph 0.962091$ & $\ph 0.996825$ & \\ \hline
 $1 $ & $\ph 0.917724$ & $\ph 0.980384 $ & $\ph 1.002270$\\ \hline
 $1.1$ & $\ph 0.895169$ & $\ph 0.967625$ & $\ph 0.992897 $\\ \hline
 $1.5$ & $\ph 0.766126$ & $\ph 0.891979$ & $\ph 0.928149 $\\ \hline
 $2 $ & $\ph 0.573703$ & $\ph 0.754224$ & $\ph 0.802883 $\\ \hline
 $2.5$ & $\ph 0.352544$ & $\ph 0.580890$ & $\ph 0.642092$\\ \hline
 $3 $ & $\ph 0.109627$ & $\ph 0.380273$ & $\ph 0.454366 $\\ \hline
 $3.5$ & $-0.150549$ & $\ph 0.157791$ & $\ph 0.245264$\\ \hline
 $4$ & $-0.424867$ & $-0.082840$ & $\ph 0.018650 $\\ \hline
 $5$ & $-1.007410$ & $-0.609771$ & $-0.476508 $\\ \hline
 $5.67 $ & & $-0.993110$ & $-0.833505 $\\ \hline
 $5.8 $ & & $$ & $-0.904766 $\\ \hline
 $6$ &   & $ $ & $-1.015560$ \\ \hline
 
\end{tabular}
\end{center}
\end{table}
 We note that the mass parameter $\mu$ decreases `slowly' with increasing $N\ge 2$ and satisfies $ma \geq \mu \geq ma/\sqrt{2}.$  In Table~1 we include values of $e_{2g}$ calculated from Eq.~(\ref{eqe2g}) for $N\rightarrow\infty$, or $\lambda \rightarrow 1$: since $e_{2g} < e_k,$ we know that the Klein--Gordon bound $e_k$ is better than the $N/2$ bound, that is to say $e_2 < e_{2g}< e_k.$  Moreover, we see from Table~1 that for large $N$, the Klein--Gordon lower bound is in fact much better than the $N/2$ bound.\medskip

The ultrarelativistic limit $m\rightarrow 0$, which is perhaps meaningful in the spinless Salpeter theory, has no useful counterpart expressed by the Klein--Gordon equation. The Salpeter Hamiltonian in this limit (and for the exponential example) becomes $H = |p| - \nu e^{-r}:$ this problem is not very different qualitatively from the corresponding Schr\"odinger problem $H = p^2 -\nu e^{-r}$, which also has a discrete eigenvalue for every $\nu > 0$ not too small.
By contrast, the Klein--Gordon equation (\ref{ekexp}) with $\mu = 0$ implies $e_k = 0$, and leads to the Schr\"odinger problem $(p^2 - \nu^2 \e^{-2r})\psi = 0,$ which has a solution only for very particular $\nu$ not too small.  Clearly, the Klein--Gordon lower bound is not useful for the ultrarelativistic limit.

\section{Supercriticality and system size}We consider first a one-body problem with an attractive (vector) potential $V(r) = vf(r)$ in three spatial dimensions.  We contrast four theories: Dirac, Klein--Gordon, Salpeter, and Schr\"odinger.  For the singular Coulomb or gravitational case $f(r) = -1/r$, all four theories agree that binding is obtained for arbitrarily small $v$.  Meanwhile, the three relativistic theories agree that the coupling $v$ cannot be very large: we have for Dirac $v < 1,$ for Klein--Gordon $v < \half,$ and for Salpeter $ v < 2/\pi.$ For these three relativistic theories, the discrete eigenvalues are all positive and the supercritical phenomenon that we discuss below does not arise. None-the-less, the coupling restrictions lead to limitations on the system size, as we have discussed earlier for a static-gravity boson-star model \cite{HL08}.\medskip

We now set aside the Coulomb case and consider bounded short-range potentials such as the exponential potential $f(r) = -e^{-r}$ discussed in the previous section.  For this class of problems, the four theories now agree that (in three dimensions) binding is obtained only for $v$ sufficiently large.  The term `critical coupling' is used for the coupling value $v_c$, in each theory, beyond which a discrete eigenvalue appears.  As $v$ is increased further, this eigenvalue decreases.  In the fully relativistic theories another critical value $v = v_s$ is reached at which $E = -m.$ This situation is called `supercriticality' and $v_s$ is the corresponding supercritical coupling. The relativistic Dirac and Klein--Gordon theories will not accommodate larger couplings than their particular supercritical values. By contrast, for the Salpeter and Schr\"odinger cases, as $v$ is increased, the eigenvalues simply decrease without lower limit.  It is now clear that if we use the Klein--Gordon eigenvalue to bound the Salpeter eigenvalue below, a limitation is introduced which was not originally present in the Salpeter problem itself, namely that the coupling should not be very large. One could perhaps argue that, just as the restriction $E > -m$, which emerges internally and technically from the theory of the Klein--Gordon equation, is `understood' in terms of another theory, namely pair creation in quantum field theory; thus, significance of this constraint could be acquired by the Salpeter theory from the Klein--Gordon theory.  All we are saying here is that, if the Salpeter theory claims to be partially  relativistic, then it is natural to impose the condition that $E > -m$, or, what is sufficient, that $v$ not be too large.  Using the Klein--Gordon theory to bound Salpeter below achieves this automatically.  In terms of the $N$-body reductions discussed in this paper, this means that the scaled coupling $\nu = \gamma^{\half}v a/2$ should not be too large.  Since $\gamma = \half N(N-1),$ this means, for a given pair interaction, that the particle number $N$, in turn,  should not be too large.
Thus we arrive at the conclusion that such relativistic systems in their ground states will not become giant balls as $N$ is increased,  but will divide into smaller aggregates.  For the exponential example with $m = 1,$ $a = \sqrt{2},$ $v$ small and $N$ large so that $\lambda \approx 1,$ we have from Table~1, 
$\nu = \gamma^{\half} v a/2 < 5.7,$ which is to say, $N < 11.4/v.$  Thus we conclude that a phenomenon which seemed to be a special feature of the $-1/r$ potential may be more widely present in relativistic many-body systems.

\section{Conclusion}The principal disadvantage of the semirelativistic Salpeter equation is that it is  not fully relativistic. It perfectly well accommodates the harmonic oscillator, a strong interaction not at all tolerated (as a vector potential) by the Klein--Gordon or Dirac theories.  Moreover, the Salpeter Hamiltonian is non-local, and this introduces technical problems for spectral estimation.  The main advantages of the Salpeter theory are (i) the ease with which the relativistic many-body problem can be formulated and (ii) the fact that the energy spectrum is characterized, and may therefore be estimated, variationally. For a wide class  of attractive pair potentials that vanish at infinity, we have shown  in this paper that Jacobi relative coordinates and the permutation symmetry of the boson states allow us to construct ground-state energy bounds in terms of certain reduced one-body problems: the lower bound is provided by the ground-state energy of a Klein--Gordon problem; and the upper bound is given by  the expectation of a Salpeter Hamiltonian with respect to a scale-optimized Gaussian trial function. These bounds may well have theoretical as well as computational applications.  For example, as $N$  increases, the effective potential coupling $\nu$ in the scaled one-body Klein--Gordon problem (\ref{kg1}) increases by the factor $\sqrt{N(N-1)}\sim N$ whereas the kinetic-energy term remains unchanged.  Even for bounded potentials, such as the exponential potential, this implies that this lower-bound Klein--Gordon problem will, for sufficiently large $N$, become supercritical, with $e < -\mu$.  This may, in turn, indicate that the $N$-body problem will not be stable for large aggregates, and will necessarily break  up into smaller fragments.  This would be the expression of a truly relativistic many-body phenomenon with no pre-image in non-relativistic theories. 
\section*{Acknowledgements}

One of us (RLH) gratefully acknowledges both partial financial support
of this research under Grant No.\ GP3438 from the Natural Sciences
and Engineering Research Council of Canada and the hospitality of
the Institute for High Energy Physics of the Austrian Academy of
Sciences, Vienna, where part of the work was done.\medskip

 \section*{Appendix~A}
 For definiteness, we consider the relative momenta $\mbox{{\boldmath $\pi$}}_i$ and $\mbox{{\boldmath $\pi$}}_j$ with $i, j >1.$ Since $[\mbox{{\boldmath $\pi$}}] = B[\mbox{{\boldmath $p$}}],$ we have
$$\mbox{{\boldmath $\pi$}}_i\cdot\mbox{{\boldmath $\pi$}}_j = \left(\sum_k B_{ik}{\bf p}_k\right)\cdot\left(\sum_k B_{jk}{\bf p}_k\right) = \sum_{k}B_{ik}B_{jk}\left({\bf p}_k^2\right) + \sum_{k\ne l}B_{ik}B_{jl}\left({\bf p}_k\cdot {\bf p}_l\right).\eqno{\rm (A1)}$$
By using the boson symmetry of the wave function we have
$$\langle \mbox{{\boldmath $\pi$}}_i\cdot\mbox{{\boldmath $\pi$}}_j\rangle =\left(\sum_{k}B_{ik}B_{jk}\right)\langle {\bf p}_1^2\rangle + \left(\sum_{k\ne l}B_{ik}B_{jl}\right)\langle {\bf p}_1\cdot{\bf p}_2\rangle.\eqno{\rm (A2)}$$
But orthogonality of the rows of $B$ to the first row tells us
$$0 = \sum_{k}B_{ik} = \left(\sum_{k}B_{ik}\right)\left(\sum_{l}B_{jl}\right) = \left(\sum_{k}B_{ik}B_{jk}\right) + \left(\sum_{k\ne l}B_{ik}B_{jl}\right).\eqno{\rm (A3)}$$
If $ i = j$, then the orthogonality of the matrix $B$ tells us that
$$\sum_{k}B_{ik}^2 = 1= -\sum_{k\ne l}B_{ik}B_{il}.$$
Meanwhile, if $i \ne j,$ we know that the corresponding rows of $B$ are orthogonal; hence, in this case, $\sum_{k}B_{ik}B_{jk} = 0;$ thus, both of the coefficients in (A2) are zero.  Hence we conclude for all $i,j > 1$
$$\langle \mbox{{\boldmath $\pi$}}_i\cdot\mbox{{\boldmath $\pi$}}_j\rangle = \delta_{ij}\langle \mbox{{\boldmath $\pi$}}_2^2\rangle = \delta_{ij}\left(\langle {\bf p}_1^2\rangle - \langle {\bf p}_1\cdot {\bf p}_2\rangle\right).\eqno{\rm (A4)}$$
The proof for the relative coordinates $\{\mbox{{\boldmath $\rho$}}_i\}$ is identical.

\end{document}